\documentclass[pra,twocolumn,showpacs,superscriptaddress,floatfix]{revtex4}
\newcount\stile
\ifnum\stile=0\def\ausilio{\immediate\openout15=\jobname.sty}\fi
\ifnum\stile=1\def\ausilio{}\fi

\font\titolo=cmbx12\font\titolone=cmbx10 scaled\magstep 2%
\font\sc=cmcsc10\font\css=cmcsc8%
\font\ottorm=cmr8%
\def\st{\scriptstyle}%
\font\tenmib=cmmib10 \font\eightmib=cmmib8
\font\sevenmib=cmmib7\font\fivemib=cmmib5
\font\ottoit=cmti8\font\fiveit=cmti5\font\sixit=cmti6
\font\fivei=cmmi5\font\sixi=cmmi6\font\ottoi=cmmi8
\font\ottorm=cmr8
\font\ottosy=cmsy8\font\sixsy=cmsy6\font\fivesy=cmsy5
\font\ottobf=cmbx8\font\sixbf=cmbx6\font\fivebf=cmbx5%
\font\ottocss=cmcsc8%

\def\ottopunti{\def\rm{\fam0\ottorm}\def\it{\fam6\ottoit}%
\def\bf{\fam7\ottobf}%
\textfont1=\ottoi\scriptfont1=\sixi\scriptscriptfont1=\fivei%
\textfont2=\ottosy\scriptfont2=\sixsy\scriptscriptfont2=\fivesy%
\textfont4=\ottocss\scriptfont4=\sc\scriptscriptfont4=\sc%
\textfont5=\eightmib\scriptfont5=\sevenmib\scriptscriptfont5=\fivemib%
\textfont6=\ottoit\scriptfont6=\sixit\scriptscriptfont6=\fiveit%
\textfont7=\ottobf\scriptfont7=\sixbf\scriptscriptfont7=\fivebf%
\setbox\strutbox=\hbox{\vrule height7pt depth2pt width0pt}%
\normalbaselineskip=9pt\rm}
\let\nota=\ottopunti%

\mathchardef\Ba   = "050B  
\mathchardef\Bb   = "050C  
\mathchardef\Bg   = "050D  
\mathchardef\Bd   = "050E  
\mathchardef\Be   = "0522  
\mathchardef\Bee  = "050F  
\mathchardef\Bz   = "0510  
\mathchardef\Bh   = "0511  
\mathchardef\Bthh = "0512  
\mathchardef\Bth  = "0523  
\mathchardef\Bi   = "0513  
\mathchardef\Bk   = "0514  
\mathchardef\Bl   = "0515  
\mathchardef\Bm   = "0516  
\mathchardef\Bn   = "0517  
\mathchardef\Bx   = "0518  
\mathchardef\Bom  = "0530  
\mathchardef\Bp   = "0519  
\mathchardef\Br   = "0525  
\mathchardef\Bro  = "051A  
\mathchardef\Bs   = "051B  
\mathchardef\Bsi  = "0526  
\mathchardef\Bt   = "051C  
\mathchardef\Bu   = "051D  
\mathchardef\Bf   = "0527  
\mathchardef\Bff  = "051E  
\mathchardef\Bch  = "051F  
\mathchardef\Bps  = "0520  
\mathchardef\Bo   = "0521  
\mathchardef\Bome = "0524  
\mathchardef\BG   = "0500  
\mathchardef\BD   = "0501  
\mathchardef\BTh  = "0502  
\mathchardef\BL   = "0503  
\mathchardef\BX   = "0504  
\mathchardef\BP   = "0505  
\mathchardef\BS   = "0506  
\mathchardef\BU   = "0507  
\mathchardef\BF   = "0508  
\mathchardef\BPs  = "0509  
\mathchardef\BO   = "050A  
\mathchardef\BDpr = "0540  
\mathchardef\Bstl = "053F  

\global\newcount\numsec\global\newcount\numapp
\global\newcount\numfor\global\newcount\numfig
\global\newcount\numsub
\def\veroparagrafo{\number\numsec}\def\veraformula{\number\numfor}
\def\veraappendice{\number\numapp}\def\verasub{\number\numsub}
\def\verafigura{\number\numfig}

\def\senondefinito#1{\expandafter\ifx\csname#1\endcsname\relax}

\def\SIA #1,#2,#3 {\senondefinito{#1#2}%
\expandafter\xdef\csname #1#2\endcsname{#3}\else
\write16{???? ma #1#2 e' gia' stato definito !!!!} \fi}

\def \Fe(#1)#2{\SIA fe,#1,#2 }
\def \Fp(#1)#2{\SIA fp,#1,#2 }
\def \Fg(#1)#2{\SIA fg,#1,#2 }

\def\Section(#1,#2){\advance\numsec by 1\numfor=1\numsub=1\numfig=1%
\SIA p,#1,{\veroparagrafo} %
\ifnum \stile=0\write15{\string\Fp (#1){\secc(#1)}}\fi%
\hbox to \hsize{\titolo\hfill \number\numsec. #2\hfill%
\expandafter{\alato(sec. #1)}}\*}

\def\Appendix(#1,#2){\advance\numapp by 1\numfor=1\numsub=1\numfig=1%
\SIA p,#1,{A\veraappendice} %
\ifnum \stile=0\write15{\string\Fp (#1){\secc(#1)}}\fi%
\hbox to \hsize{\titolo Appendix A\number\numapp. #2\hfill%
\expandafter{\alato(app. #1)}}%
\*%
}

\def\etichetta(#1){(\veroparagrafo.\veraformula)%
\SIA e,#1,(\veroparagrafo.\veraformula) %
\global\advance\numfor by 1%
\ifnum \stile=0\write15{\string\Fe (#1){\equ(#1)}}\fi%
}

\def\etichettaa(#1){(A\veraappendice.\veraformula)%
\SIA e,#1,(A\veraappendice.\veraformula) %
\global\advance\numfor by 1%
\ifnum \stile=0\write15{\string\Fe (#1){\equ(#1)}}\fi%
}

\def\getichetta(#1){\veroparagrafo.\verafigura%
\SIA g,#1,{\veroparagrafo.\verafigura} %
\global\advance\numfig by 1%
\ifnum \stile=0\write15{\string\Fg (#1){\graf(#1)}}\fi%
}

\def\etichettap(#1){\veroparagrafo.\verasub%
\SIA p,#1,{\veroparagrafo.\verasub} %
\global\advance\numsub by 1%
\ifnum \stile=0\write15{\string\Fp (#1){\secc(#1)}}\fi%
}

\def\Eq(#1){\eqno{\etichetta(#1)\alato(#1)}}
\def\eq(#1){\etichetta(#1)\alato(#1)}
\def\Eqa(#1){\eqno{\etichettaa(#1)\alato(#1)}}
\def\eqa(#1){\etichettaa(#1)\alato(#1)}
\def\eqg(#1){\getichetta(#1)\alato(fig. #1)}
\def\sub(#1){\0\palato(p. #1){\bf \etichettap(#1).}}
\def\asub(#1){\0\palato(p. #1){\bf \etichettapa(#1).}}
\def\apprif(#1){\senondefinito{e#1}%
\eqv(#1)\else\csname e#1\endcsname\fi}

\def\equv(#1){\senondefinito{fe#1}$\clubsuit$#1%
\write16{eq. #1 non e' (ancora) definita}%
\else\csname fe#1\endcsname\fi}
\def\grafv(#1){\senondefinito{fg#1}$\clubsuit$#1%
\write16{fig. #1 non e' (ancora) definito}%
\else\csname fg#1\endcsname\fi}
\def\secv(#1){\senondefinito{fp#1}$\clubsuit$#1%
\write16{par. #1 non e' (ancora) definito}%
\else\csname fp#1\endcsname\fi}

\def\eqo{{\global\advance\numfor by 1}}
\def\equ(#1){\senondefinito{e#1}\equv(#1)\else\csname e#1\endcsname\fi}
\def\graf(#1){\senondefinito{g#1}\grafv(#1)\else\csname g#1\endcsname\fi}
\def\figura(#1){{\css Figura} \getichetta(#1)}
\def\secc(#1){\senondefinito{p#1}\secv(#1)\else\csname p#1\endcsname\fi}
\def\sec(#1){{\secc(#1)}}
\def\refe(#1){{[\secc(#1)]}}

\def\BOZZA{
\def\alato(##1){\rlap{\kern-\hsize\kern-.5truecm{$\scriptstyle##1$}}}
\def\palato(##1){\rlap{\kern-.5truecm{$\scriptstyle##1$}}}
}

\def\alato(#1){}
\def\galato(#1){}
\def\palato(#1){}

{\count255=\time\divide\count255 by 60 \xdef\hourmin{\number\count255}
        \multiply\count255 by-60\advance\count255 by\time
   \xdef\hourmin{\hourmin:\ifnum\count255<10 0\fi\the\count255}}

\def\oramin{\hourmin }

\def\data{\number\day/\ifcase\month\or gennaio \or febbraio \or marzo \or
aprile \or maggio \or giugno \or luglio \or agosto \or settembre
\or ottobre \or novembre \or dicembre \fi/\number\year;\ \oramin}
\setbox200\hbox{$\scriptscriptstyle \data $}

\def\fiat{}
\let\a=\alpha \let\b=\beta      \let\e=\varepsilon
\let\z=\zeta        
\let\m=\mu    \let\n=\nu         \let\p=\pi     
\let\s=\sigma \let\t=\tau   \let\f=\varphi 
      
 \let\D=\Delta   
    \let\Si=\Sigma\let\F=\Phi    \let\Ps=\Psi

\def\CC{{\cal C}}
\def\FF{{\cal F}}

\let\ig=\int
\let\io=\infty
\let\==\equiv
\let\0=\noindent

\def\\{\hfill\break}
\def\lis#1{\overline#1}

\def\*{\vskip3mm}
\def\ie{{\it i.e. }}
\def\eg{{\it e.g. }}

\let\dpr=\partial

\def\defi{\,{\buildrel def\over=}\,}

\def\V#1{{\bf#1}}
\def\media#1{{\langle#1\rangle}}
\def\fra#1#2{{#1\over#2}}
\def\crcl{\,\raise.5mm\hbox{$\st\rm o$}\,}%
\def\otto{\,{\kern-1.truept\leftarrow\kern-5.truept\to\kern-1.truept}\,}
\def\tende#1{\,\vtop{\ialign{##\crcr\rightarrowfill\crcr
 \noalign{\kern-1pt\nointerlineskip} \hskip3.pt${\scriptstyle
 #1}$\hskip3.pt\crcr}}\,}
\newdimen\xshift \newdimen\xwidth \newdimen\yshift \newdimen\ywidth

\def\ins#1#2#3{\vbox to0pt{\kern-#2\hbox{\kern#1 #3}\vss}\nointerlineskip}

\def\eqfig#1#2#3#4#5{
\par\xwidth=#1 \xshift=\hsize \advance\xshift
by-\xwidth \divide\xshift by 2
\yshift=#2 \divide\yshift by 2%
{\hglue\xshift \vbox to #2{\vfil
#3 \includegraphics{#4.eps}
}\hfill\raise\yshift\hbox{#5}}}

\def\8{\write12}

\input #1.sty \closein13\fi

\newcommand\revtex{{R\kern-0.4mm\lower0.5mm\hbox{E}\kern-0.4mm V\kern-0.3mm%
\lower0.5mm\hbox{T}\kern-0.4mm E\kern-.3mm \lower0.5mm\hbox{X}}}

\usepackage{eqalignno}
\begin{document}
\ausilio

\fiat
\voffset0.5truecm

\centerline{\titolone Entropy, Thermostats and}

\centerline{\titolone Chaotic Hypothesis}
\*
\centerline{\bf Giovanni Gallavotti}

\centerline{Fisica and I.N.F.N. Roma 1}
\centerline{
\today
}
\*\*
\0{\bf Abstract: \it The chaotic hypothesis is proposed as a basis
for a general theory of nonequilibrium stationary states.}
\*

\0{\bf 1. \it Stationary states and thermostats.}
\numsec=1\numfor=1\*

The problem is to develop methods to establish relations between time
averages of a few observables associated with a system of particles
subject to work-performing external forces and to thermostat-forces
that keep the energy from building up, so that it can be considered in
a stationary state.

The stationary state will correspond to a probability distribution on
phase space $\FF$ so that

$$\eqalign{
&\fra1\t\sum_{j=0}^{\t-1} F(S^jx)\tende{\t\to\io} \ig_\FF
F(y)\,\m(dy),\qquad
{\rm or}\cr
&\fra1\t\ig_0^\t F(S_tx)\,dt\tende{\t\to\io} \ig_\FF
F(y)\,\m(dy)\cr}\Eq(e1.1)$$
for all $x$ but a set of zero volume: the first refers to cases in
which dynamics is a map $S:\FF\to\FF$ and the second when it is a flow
defined by a differential equation on $\FF$:

$$\dot{ x}={\bf f}_{\V E}({x})\Eq(e1.2)$$
where ${\V f}_{\V E}$ contains internal forces, external
forces depending on a few parameters $\V E=(E_1,\ldots,E_n)$, and
thermostats forces. In general the divergence

$$\s(x)=-\sum_j \dpr_{x_j} f_{\V E,j}(x)\Eq(e1.3)$$
is not zero, except in absence of external forces $\V E$ and
of thermostat forces (\ie in the equilibrium case).

A fairly realistic example is the following:

\eqfig{110pt}{90pt}{}{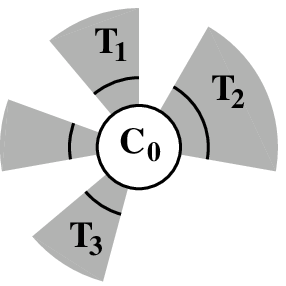}{}

\0{\nota Fig.1 ``Thermostats'', or reservoirs, occupy finite regions
outside $C_0$, \eg sectors $C'_a\subset R^3$, $a=1,2\ldots$, marked
$T_a$ located beyond ``buffers'' $\CC_a$: the buffers (representing a
the {\it walls} separating the system from the thermostats) simply
have their boundaries marked. The reservoir particles are constrained
to have a {\it total} kinetic energy $K_a$ constant, by suitable
forces $\Bth_a$, so that their ``temperatures'' $T_a$, see (1.5),
are well defined, \cite{Ga06}. Buffers and reservoirs have
{\it arbitrary sizes}.\vfil} \*

The system contains $N_0$ particles in a configuration $\V X_0$
contained in $\CC_0$ and $N_i,N'_i$ particles in configurations that
will be denoted $\V X_i,\V X'_i$ contained in the buffer regions
$\CC_i$, henceforth called {\it wall}. and in the thermostat regions
$\CC'_i$, $i=1,\ldots,n$, respectively. The equations of motion are,
for $i=0$ and $i>0$ respectively,

$$\eqalignno{
&\ddot {\V X}_0=-\dpr_{\V X_0} (U_0(\V X_0)+{\sum}_{i>0}
W_{0i}(\V X_0,\V X_i))+ \V E(\V X_0)\cr
&\ddot {\V X}_i=-\dpr_{\V X_i} (U_i(\V X_i)+
W_{0i}(\V X_0,\V X_i)+ W_{i,i'}(\V X_i,\V X_{i'}))\cr
&\ddot {\V X}'_{i}=-\dpr_{\V X'_i} (U'_i(\V X'_i)+
W_{i,i'}(\V X_i,\V X_{i}'))-\a_i\,\dot{\V X}'_i&
\eq(e1.4)\cr}$$
where $U_i,U'_i$ are the interaction energies for the particles in
$\CC_i$, $i=0,1,\ldots,n$ and in $\CC'_i$, $i=1,\ldots,n$; $\V E(\V
X_0)$ is the external force working on the system in $\CC_0$ and $-\a_i
\dot{\V X}'_i$ is the {\it thermostat force}: which is the force
prescribed by {\it Gauss' principle of least effort}, see Appendix
A9.4 in \cite{Ga00}, to impose the contraints ($k_B\=$
Boltzmann's constant)

$$\fra12 \dot{\V X'}_i^2=\fra{d}2 N'_i k_B T_i\Eq(e1.5)$$
which gives, after a simple application of the principle,

$$\a_i= \fra{L_i-\dot U_i'}{N'_i\,k_B T_i}\Eq(e1.6)$$
where $L_i$ is the work done per unit time by the particles $\V
X_i\in\CC_i$  on those in $\V X_i'\in\CC'_i$, \ie on the thermostats.

Other thermostat models could be considered: however their particular
structure should not influence the statistical properties of the
particles in $\CC_0$. In particular I think that replacing the
container $\CC'_i$ with an {\it infinite} container in which particles
are initially in a state that is an equilibrium Gibbs state at
temperature $T_i$ should lead to the same results: this is a conjecture
whose proof seems quite far at the moment.

In the following we shall regard the equations \equ(e1.4) as first
order equations on the phase space coordinates $x\=\{\dot{\V X}_i,\V
X_i\}_{i=0}^n$. As such the equations do not conserve volume of
phase space: in fact the divergence of the equations in this space
is $-\s(x)$ with

$$\eqalignno{
\s(x)=&\sum_{i>0} \fra{L_i}{k_B T_i}\fra{d N'_i-1}{d N'_i}
-\sum_{i>0}\fra{\dot U'_i}{k_B T_i} \fra{d N'_i-1}{d N'_i}=\cr
=&\sum_{i>0} \fra{L_i}{k_B T_i}\fra{d N'_i-1}{d N'_i}+\dot\F&  \eq(e1.7)\cr}$$
where $\F\defi-\sum_{i>0}\fra{U'_i}{k_B T_i}$, as it can be checked by
direct computation.

Since $L_i=-\dot{ X}'_i\cdot\dpr_{X'_i} W_{i,i'}\= +\dot {
X}_i\cdot\dpr_{X_i} W_{i,i'}-\dot W_{i,i'}$ and the expression
\equ(e1.7) is the sum over $i>0$ of
$-\fra{d}{dt}\Big(\fra12\dot{X}_i^2+U_i\Big)-\dot { X}_i\dpr_{X_i}
W_{i,0}$ which has the form $\dot\Ps_i+Q_i$ where $Q_i$ is the work
per unit time done by the forces due to particles in $\CC_0$ on the
particles in $\CC_i$: we identify therefore $Q_i$ with the {\it heat}
generated per unit time by the forces acting on $\CC_0$ and transfered
first to the walls $\CC_i$ and, subsequently, to the thermostats in
$\CC_i'$.

Thus setting $\e(x)\defi \sum_{i>0}\fra{Q_i}{k_B T_i}$ it is
(for notational simplicity, and keeping in mind that $N'_i$
should be thought as large,  we shall neglect $O({N_i'}^{-1})$)

$$\s(x)=\e(x)+\dot R\Eq(e1.8)$$
where $R(x)=-\sum_{i}\fra{W_{i,i'}+U'_i+U_i +\fra12\dot{X_i}^2}{k_B T_i}$.

\* \0{\it Remark:} (1) In this model, as well as in a large number of
others, one has therefore the natural interpretation of $\s(x)$ as the
{\it entropy creation} per unit time: this is because for large time
the average of the l.h.s., $\s(x)$, over a time interval and the
corresponding average of $\e(x)=\sum_{i>0}\fra{Q_i}{k_B T_i}$ become
equal at large time because they differ by $\fra1\t(R(S_tx)-R(x))$, at
least if $R$ is bounded, as it is convenient to suppose for
simplicity. This is a strong assumption but it will not be discussed
here: it has to do with the problem of thermostats ``efficiency'' and
its violation may lead to interesting consequences, see
\cite{Ga06,GG06}.
\\ 
(2) It should be noted that the walls $\CC_i$ could be missing and the
particles in $\CC_0$ be directly in contact with the thermostats: in
this case there will be no $W_{i,i'}$ but instead there will be
potentials $W_{0,i'}$: the analysis would be entirely analogous with
$\fra{Q_i}{k_B T_i}$ replaced by $\fra{Q'_{i'}}{k_B T_i}$ with
$Q'_{i'}$ being the work per unit time done by the particles in
$\CC_0$ on the thermostat particles in $\CC'_i$ and $R=-\sum_{i}
\fra{W_{i,i'}+U'_i}{k_B T_i}$. In this case if the porentials of
interactions are bounded the $R$ will be also bounded without any
extra assumption.
\\ 
(3) The $L_i$ in Eq.\equ(e1.7) is the work ceded by the walls to
thermostats: therefore it can be interpreted as the heat $Q'_i$ ceded
by the paricles in $\CC_i$ to the thermostat in $\CC'_i$: hence the
alternatice representation $\s(x)=\e'(x)+\dot\F$, \equ(e1.7), is
possible with $\e'(x)=\sum_{i>0}\fra{Q'_i}{k_B T_i}$. Also in this case the
remainder $\F$ is bounded if the interaction potentials are bounded
and the discussion that follows applies to both $\e(x)$ and $\e'(x)$,
which are thus equivalent for the purpose of fluctuation analysis.

\*
\0{\bf2. \it The hypothesis.}
\numsec=2\numfor=1\*

\0{\bf Chaotic Hypothesis: \it Motions developing on the
attracting set of a chaotic system can be regarded as a transitive
hyperbolic system.}
\*

A general result is that transitive hyperbolic systems have the
property \equ(e1.1), with $\m$ a uniquely determined probability
distribution on phase space, \cite{BR75}.

Of course a flow can be studied via a {\it Poincar\'e map} $S$ defined
by a {\it timing event} $\Si$. The latter is defined by a surface in
phase space which is crossed by all trajectories infinitely many times
(typically $\Si$ is the union of a few connected surface elements
$\Si=\cup_i \Si_i$, but in general it is {\it not} connected: \ie it is
a finite collection of connected pieces). The timing event occurs when
a trajectory crosses $\Si$ at a point $x$ and time $t_0$: and $S$ maps
it into the next timing event $Sx$ occurring, at some time $t_1$, on
the trajectory $t\to S_tx$: hence $x'=Sx\defi S_{t_1-t_0}x\in \Si$.

For model \equ(e1.4) there is a direct relation
between $\s(x)$, $x\in \Si$, and the Jacobian determinant
$\det\dpr_xS(x)$; setting $R(t_1)\=R(S_{t_1-t_0}x),\,R(t_0)\=R(x)$,
it is

$$\eqalign{
&-\log |\det\dpr_xS(x)|=\ig_{t_0}^{t_1} \s(S_t x)\,dt=\cr
&=\ig_{t_0}^{t_1}
\e(S_tx)\,dt+ R(t_1)-R(t_0)=\cr
&=\sum_{i>0}\fra{\ig_{t_0}^{t_1}Q_i\,dt}{k_B T_i}+ R(t_1)-R(t_0)\cr
}\Eq(e2.1)$$
The theory of evolutions described by flows or described by maps are
therefore very closely related as the above remarks show, at least for
what concerns the analysis of the entropy creation rate and its
fluctuations.

The second viewpoint should be taken whenever $\s(x)$ has
singularities: which can happen if the interaction potentials are
unbounded (\eg of Lennard-Jones type) or if the thermostats sizes tend
to infinity, see \cite{BGGZ05}.
\*

\0{\it3. Dimensionless entropy and fluctuation theorem.}
\numsec=3\numfor=1\*

Interesting properties to study are related to the fluctuations of
{\it entropy creation} averages.  Restricting the analysis to the
model \equ(e1.4), define the {\it entropy creation} rate to be

$$\e_+=\lim_{\t\to\io} \fra1\t\ig_0^\t
\s(S_tx)\,dt
=\lim_{\t\to\io} \fra1\t\ig_0^\t \e(S_tx)\,dt\Eq(e3.1)$$
by the remark at the end of Sec.1.

Assuming that the system is {\it dissipative}, which by definition will
mean $\e_+>0$, consider the random variable

$$p\defi\fra1\t\ig_0^\t\fra{\e(S_tx)}{\e_+}\,dt\Eq(e3.2)$$
that will be called the dimensionless phase space contraction and
considered with the distribution inherited from the SRB-distribution
$\m$ of the system.

A general property of random variables of the form $a=\fra1\t\ig_0^\t
\,F(S_tx)\,dt$, which are time averages over a time $\t$ of a smooth
observable $F$, is that, if motions are transitive and hyperbolic, the
SRB-probability distribution $\m$ that $a$ is in a closed interval
$\D$ has the form

$$P_\m(a\in\D)=\exp  (\t\,\max_{a\in\D} \z_F(a)+O(1))\Eq(e3.3)$$
for $\D\subset (a_-,a_+)$, where $a_\pm$ are two suitable values
within which the function $\z_F(a)$ is defined, analytic and convex;
the {\it fluctuation interval} $[a_-,a_+]$ contains the $\m$--average
value of $F$ and if $\D\cap[a_-,a_+]=\emptyset$ the probability
$P_\m(a\in\D)$ tends to $0$ as $\t\to\io$ faster than
exponentially. For this reason the function $\z_F(a)$ can be naturally
defined also for $a\not\in[a_-,a_+]$ by giving it the value $-\io$,
\cite{Si72,BR75,Si77,Si94}. Finally $O(1)$ means a quantity which is
bounded as $\t\to\io$ at $\D$ fixed.

The function $\z_F(a)$ is called the {\it large deviations rate} for
the fluctuations of the observable $F$.

If the motions are also {\it reversible}, \ie if there is an isometry
$I$ of phase space such that $IS_t=S_{-t}I$ or $IS=S^{-1}I$, in the
case of time evolution maps, any observable $F$ which is odd under
time reversal, \ie $F(Ix)=-F(x)$ will have a fluctuation interval
$[-a^*,a^*]$ symmetric around the origin (and containing the
SRB--average $\lis a$ of $F$).

In the case of the model \equ(e1.4) time reversibility corresponds to
the velocity inversion and the evolution is reversible in the just
defined sense. The fluctuation interval of $\s(x)/\e_+$ and of
$\e(x)/\e_+$ is therefore symmetric around the origin and $p^*\ge1$
because the averages of the two observables are $1$ by definition, see
\equ(e3.1),\equ(e3.2).

A general theorem that holds for transitive, hyperbolic motions is the
following

\* \0{\bf Fluctuation theorem: \it Given a hyperbolic, transitive and
reversible system assume that the SRB average $\s_+$ of the phase
space contraction $\s(x)$, \ie that the divergence of the equations of
motion \equ(e1.3), is $\s_+>0$. Consider the dimensionless phase space
contraction $\s(x)/\s_+$: this is an observable which has a large
deviations rate $\z(p)$ defined in a symmetric interval $(-p^*,p^*)$
and satisfying there

$$\z(-p)=\z(p)-p\s_+\Eq(e3.4)$$
}
\*

\0{\it Remarks:}
(i) The \equ(e3.4) can be regarded as valid for all $p$'s
if we follow the mentioned convention of defining $\z(p)=-\io$ for
$p\not\in[-p^*,p^*]$.
\\
(ii) By the chaotic hypothesis, abridged CH, it follows that a
relation like \equ(e3.4) should hold for the SRB distribution of the
dimensionless phase space contraction of any reversible chaotic motion
with a dense attractor or, more generally, for dimensionless phase
space contraction of the motions restricted to the attracting set, if a
time reversal symmetry holds on the motions restricted to the
attracting set, \cite{BGG97,BG97}.  Of course this is not a theorem
(mainly because hyperbolicity is a hypothesis) but it should
nevertheless apply to many interesting cases.
\\
(iii)
In particular it should apply to the model \equ(e1.4): actually in this
case it has already been remarked that the observable $\s(x)/\s_+$ and
the {\it dimensionless entropy creation rate} $\e(x)/\e_+$ have the
{\it same large deviations function}; hence \equ(e3.4) should hold for
the rate function of $p=\fra1\t\ig_0^\t\sum_{a>0} \fra{Q_a}{k_B T_a\,
  \e_+}\,dt$:

$$\z(-p)=\z(p)-p\e_+,\qquad p\in(-p^*,p^*)\Eq(e3.5)$$
\\
(iv)
The latter remark is interesting because the quantity $\e(x)\defi\sum_{a>0}
\fra{Q_a}{k_B T_a}$ has a physical meaning and can be measured in
experiments like the one described in Fig.1 or in experiments for
which there is not an obvious equation of motion (\ie no obvious
model).
\\
(v) Therefore in applications the relation \equ(e3.5) is
expected to hold quite generally and, in the general cases, it is called
{\it fluctuation relation}, abridged FR, to distinguish it from the
Fluctuation Theorem.
\\
(vi) Furthermore the quantity $\e(x)$ is a {\it local} quantity as it
depends only on the microscopic configurations of the system $\CC_0$
and of the walls $\CC_i$ in the immediate vicinity of their
separating boundary. In particular the relation \equ(e3.5) does not
depend on what happens in the bulk of the walls $\CC_i$ or on the
size of the thermostats $\CC'_i$: hence the latter can be taken to
infinity. One can also imagine that \equ(e3.5) remains valid in the
case of infinite thermostats whose particles are initially distributed
so that their emprical distribution is asymptotically a Gibbs state at
temperature $T_a$.
\\
(vii) The last few comments suggest quite a few tests of the chaotic
hypothesis and of the corresponding fluctuation relation in various
cases, see for instance \cite{BGGZ06}. Therefore the fluctuation
relation, first suggested by the simulation in \cite{ECM93}, where it has
been discovered in an experiment motivated to test ideas emerging from the
SRB theory, and subsequently proved as a theorem for Anosov systems in
\cite{GC95,Ge98}, gave rise to the chaotic hypothesis and at the
moment experiments are being designed to test its predictions.
\\
(viii) The theorem will be referred as FT. It is often written in the
form, see \equ(e3.3),\equ(e3.4),

$$\lim_{\t\to\io} \fra1\t \log \fra{P_\m(p\in\D)}{P_\m(p\in\D)}=\,
\s_+\,\max_{p\in\D}p\Eq(e3.6)$$
for $\D\subset (-p^*,p^*)$
or in the more suggestive, although slightly imprecise, form:

$$\lim_{\t\to\io} \fra1\t \log \fra{P_\m(p)}{P_\m(-p)}\,=\,p\,\s_+\Eq(e3.7)$$
which can be regarded valid for $p\in(-p^*,p^*)$.
\\
(ix) It is natural to think that the special way in which the
thermostats are implemented is not important as long as the notion of
temperature of the thermostats is clearly understood. For instance an
alternative thermostat could be a stochastic one with particles
bouncing off the walls with a Maxwellian velovity distribution at
temperature depending on the wall hit. In this context the experiment
in \cite{BCL98} appears to give an interesting confirmation.
\*

\0{\bf4. Extending Onsager-Machlup's fluctuations theory}
\numsec=4\numfor=1\*

A remarkable theory on nonequilibrium fluctuations has been started by
Onsager and Machlup, \cite{OM53a,OM53b}, and concerns fluctuations
near equilibrium and, in fact, it only deals with properties of
derivatives with respect to the external forces parameters $\V E$ {\it
evaluated} at $\V E=\V 0$.

The object of the analysis are {\it fluctuation patterns}: the
question is which is the probability that the successive values of
$F(S_t x)$ follow, for $t\in[-\t,\t]$, a preassigned sequence of values,
that I call {\it pattern} $\f(t)$, \cite{Ga97}.

In a reversible hyperbolic and transitive system consider $n$
observables $F_1,\ldots,F_n$ which have a well defined parity under
time reversal $F_j(Ix)=\pm F_j(x)$. Given $n$ functions $\f_j(t)$,
$j=1,\ldots,n$, defined for $t\in[-\fra\t2,\fra\t2]$ the question is:
which is the probability that $F_j(S_tx)\sim \f_j(t)$ for $t\in
[-\fra\t2,\fra\t2]$? the following {\it FPT theorem} gives an answer:
\*

\0{\bf Fluctuation Patterns Theorem: \it Under the assumptions of the
fluctuation theorem given $F_j,\f_j$, and given $\e>0$ and an interval
$\D\subset(-p^*,p^*)$ the joint probability with respect to the SRB
distribution

$$\eqalign{
&\fra{P_\m(|F_j(S_tx)-\f_j(t)|_{j=1,\ldots,n}<\e, p\in\D)}
{P_\m(|F_j(S_tx)\mp\f_j(-t)|_{j=1,\ldots,n}<\e, -p\in\D)}=\cr
&=\exp(\t
  \max_{p\in\D}\,p\,\s_++O(1))\cr}\Eq(e4.1)$$
where the sign choice $\mp$ is opposite to the parity of $F_j$ and
$p\defi \fra1\t\ig_{-\fra\t2}^{\fra\t2} \fra{\s(S_tx)}{\s_+}\,dt$.}
\*

\0{\it Remarks:} (i) The FPT theorem means that ``all that has to be
done to change the time arrow is to change the sign of the entropy
production'', \ie the {\it time reversed processes occur with equal
likelyhood as the direct processes if conditioned to the opposite
entropy creation}. This is made clearer by rewriting the above
equation in terms of probabilities {\it conditioned on a preassigned value
of $p$}; in fact up to $e^{O(1)}$ it becomes, \cite{Ga00}, for $|p|<p^*$:

$$\eqalign{
&\fra{P_\m(|F_j(S_tx)-\f_j(t)|_{j=1,\ldots,n}<\e,\, |\, p)}
{P_\m(|F_j(S_tx)\mp\f_j(-t)|_{j=1,\ldots,n}<\e, \, |\, -p)}=1\cr
}\Eq(e4.2)$$

\\
(ii) An immediate consequence is that defining $f_i$ the averages
$f_i\defi\fra1\t\ig_{-\fra\t2}^{\fra\t2} F_j(S_tx)$ then the SRB
probability that $f_1,\ldots,f_n$ occur in presence of an entropy
creation rate $p$ is related to the occurrence of $\mp f_1,\ldots,\mp
f_n$ in presence of the opposite entropy creation rate:
in a slightly imprecise form, see remark (viii) in Sec.3 and
\equ(e3.7), this means that

$$\lim_{\t\to\io}\fra1\t \log \fra{P_\m (f_1,\ldots, f_n,p)}{P_\m (\mp
  f_1,\ldots, \mp  f_n,-p)}= p\,\s_+.\Eq(e4.3)$$
\\
(iii) In particular if $F_j$ are odd under time reversal and
$p$ can be expressed as an (obviously odd) function of $f_1,\ldots,f_n$:
$p=\p(f_1,\ldots,f_n)$ the \equ(e4.3) can be written, \cite{Ga97},

$$\lim_{\t\to\io}\fra1\t
\log \fra{P_\m (f_1,\ldots, f_n)}{P_\m (-f_1,\ldots, -f_n)}=\,
\p(f_1,\ldots,f_n)\,\s_+\Eq(e4.4)$$
for $\p(f_1,\ldots,f_n)\in(-p^*,p^*)$: a particular case of this
relation is relevant for Kraichnan's theory of turbulence,
\cite{CDG06}.
\\
(iv) An interesting application, \cite{Ga96a,Ga96}, of\equ(e4.3) with
$j_j(x)=\dpr_{E_j}
\s(x)$ is that, setting $J_j=\m(j_j)\=\media{j_i}_\m$, it is

$$L_{jk}=\dpr_{E_k} J_j|_{\V E=\V0}=L_{kj}\Eq(e4.5)$$
Since in several interesting cases $J_j$ have the interpretation
of ``thermodynamic currents'' (\ie currents divided by $k_B T$ if $T$
is the temperature) generated by the ``thermodynamic forces'' $E_j$
the \equ(e4.5) have the interpretation of {\it Onsager reciprocal
relations}. In fact also the expressions

$$L_{jk}=\fra12\ig_{-\io}^\io \m(\s_k(S_tx)\s_j(x))_{\V E=\V0} \,dt\Eq(e4.6)$$
follow from FPT and have the interpretation of {\it Green--Kubo
formulae}. The above relations have been derived under the extra
simplification that $\s\=0$ for $\V E=\V 0$ which is satisfied in
several cases, see comment following Eq.(3.5) in \cite{Ga96}. However
what is really necessary is that $\media{\s}_{\V E=\V0}=\V 0$, which
is an even weaker assumption because the analysis in \cite{Ga96a} is,
{\it verbatim}, unchanged if instead of $\s=0$ for $\V E=\V0$ one has
$\media{\s}_{\V E=\V0}=\V 0$.
\\
(v) The assumption of reversibility at $\V E\ne\V0$, which is necessary
for the FPT, is not really necessary to derive \equ(e4.6) (hence
\equ(e4.5)) as shown in \cite{GR97} where such relations are derived
under the only assumption that for just $\V E=\V 0$ the motions is
reversible.
\\
(vi) A further application of FPT is its relation with the theory of
intermittency, see \cite{Ga00a,Ga01a}.
\\
(vii) The above analysis and the arbitrariness of the walls
$\CC_i$ hints that even if the thermostating mechanism is quite
different, for instance it is generated by viscous forces $-\n_i
\dot{\V X}_i$ hence not reversible, nevertheless the quantity $\e(x)$
will satisfy a FR.
\\
(viii) In any event it appears that the total phase space
divergence $\s(x)$ is not directly physically relevant and in fact
{\it it is not physically meaninglful}. Since it differs from the
physically measurable entropy production $\e(x)$ by a total derivative it can
only be used to infer properties of the latter, as done in the FR: of
course a FR will hold also for $s(x)$ in the reversible cases. However
given the possibly very large (arbitrarily large) size of the contributions
to $\s(x)$ due to the total derivative $\dot R(x)$ to \equ(e2.1), or
\equ(e1.8), the time scale for the large fluctuations of
$p'=\fra1\t\ig_0^\t \fra{\s(S_tx)}{\e_+}$ easily becomes unobservably
large while the time scale for the fluctuations of $\e(x)$ remains
independent on the size of the walls $\CC_i$ and of the thermostats
$\CC'_i$.
\*

\0{\bf5. JF, BF and fluctuation relations}
\numsec=5\numfor=1\*

An immediate consequence of FT is that

$$\media{ e^{-\ig_0^\t \e(S_tx)\,dt}}_{SRB}=e^{O(1)}
\Eq(e5.1)$$
\ie $\media{ e^{-\ig_0^\t \e(S_tx)\,dt}}_{SRB}$ stays bounded as
$\t\to\io$.  This is a relation that I will call {\it Bonetto's
formula} and denote it BF, see Eq.(9.10.4) in \cite{Ga00}; it can be
also written, somewhat imprecisely and for mnemonic purposes,
\cite{Ga98b},

$$\media{ e^{-\ig_0^\t
\e(S_tx)\,dt}}_{SRB}\tende{\t\to\io}1\Eq(e5.2)$$
which {\it would be exact} if the  FT in the form \equ(e3.7) held for
finite $\t$ (rather than in the limit as $\t\to\io$).

This relation bears resemblance to {\it Jarzinsky's formula},
henceforth JF, which deals with a canonical Gibbs distribution (in a
finite volume) corresponding to a Hamiltonian $H_0(p,q)$ and
temperatute $T=(k_B \b)^{-1}$, and with a time dependent family of
Hamiltonians $H(p,q,t)$ which interpolates between $H_0$ and a second
Hamiltonian $H_1$ as $t$ grows from $0$ to $1$ (in suitable units)
which is called {\it a protocol}.

Imagine to extract samples $(p,q)$ with a canonical probability
distribution $\m_0(dpdq)= Z_0^{-1}e^{-\b H_0(p,q)}dpdq$, with $Z_0$
being the canonical partition function, and let $S_{0,t}(p,q)$ be the
solution of the Hamiltonian {\it time dependent} equations $\dot
p=-\dpr_q H(p,q,t),\dot q=\dpr_p H(p,q,t)$ for $0\le t\le1$. Then
JF, \cite{Ja97,Ja99}, gives:

\* \0{\it Let $(p',q')\defi S_{0,1}(p,q)$
and let $W(p',q')\defi H_1(p',q')-H_0(p,q)$, then the
distribution $Z_1^{-1} e^{-\b H_1(p',q')}dp'dq'$ is
exactly equal to $\fra{Z_0}{Z_1} e^{-\b W(p',q')}\m_0(dp dq)$. Hence

$$\media{e^{-\b W}}_{\m_0}=\fra{Z_1}{Z_0}=e^{-\b \D F(\b)}\Eq(e5.3)$$ 
where the average is with respect to the Gibbs distribution $\m_0$ and
$\D F$ is the free energy variation between the equilibrium states
with Hamiltonians $H_1$ and $H_0$ respectively.}
\*
%

\0{\it Remark:} (i) The reader will recognize in this {\it exact
identity} an instance of the Monte Carlo method. Its interest lies in
the fact that it can be implemented {\it without actually knowing}
neither $H_0$ nor $H_1$ nor the {\it protocol} $H(p,q,t)$. If one
wants to evaluate the difference in free energy bewteen two
equilibrium states at the same temperature of a system that one can
construct in a laboratory then ``all one has to do'' is 
\\ 
(a) to fix a protocol, \ie a procedure to transform the forces acting
on the system along a well defined {\it fixed once and for all} path
from the initial values to the final values in a fixed time interval
($t=1$ in some units), and 
\\ 
(b) measure the energy variation $W$ generated by the machines
implementing the protocol. This is a really measurable quantity at
least in the cases in which $W$ can be interpreted as the work done on
the system, or related to it.

Then average of the exponential of $-\b W$ with respect to a large
number of repetition of the protocol. This can be useful even, and
perhaps mainly, in biological experiments.
\\
(ii) If the ``protocol'' conserves energy (like a Joule expansion of a
gas) or if the difference $W=H_1(p',q')-H_0(p,q)$ has zero average in the
equilibrium state $\m_0$ we get, by Jensen's inequality (\ie by the
convexity of the exponential function $\media{e^A}\ge e^{\media A}$),
that $\D F\le0$ as expected from Thermodynamics.
\\
(iii) The measurability of $W$ is a difficult question, to be
discussed on a case by case basis. It is often possible to
identify it with the ``work done by the machines implementing the
protocol''.
\*

The two formulae \equ(e5.2) and \equ(e5.3) are however quite different:

(1) the $\ig_0^\t \s(S_tx)\, dt$ is an entropy creation rather than
the energy variation $W$.

(2) the average is over the SRB distribution of a stationary state, in general
    out of equilibrium, rather than on a canonical equilibrium state.

(3) the BF says that $\media{ e^{-\ig_0^\t \e(S_tx)\,dt}}_{SRB}$ is
bounded, \equ(e5.1), as $\t\to\io$ rather than being $1$
exactly. However a careful analysis of the meaning of $W$ would lead
to concluded that also JF necessitates corrections, particularly in
thermostatted systems, \cite{Ja97}.
\*

The JF has proved useful in various equilibrium problems (to
evaluate the free energy variation when an equilibrium state with
Hamiltonian $H_0$ is compared to one with Hamiltonian $H_1$); hence it has
some interest to investigate whether \equ(e5.2) can have some
consequences.

If a system is in a steady state and produces entropy at rate $\e_+$
(\eg a living organism feeding on a background) the FT \equ(e3.4) and
is consequence BF, \equ(e5.2), gives us informations on the the
fluctuations of entropy production, \ie of heat produced, and
\equ(e5.2) {\it could be useful}, for instance, to check that all
relevant heat transfers have been properly taken into account.

\* \0{\bf Acknowledgements:} 
{\it Source of a talk at the Durham Symposium ``Dynamical Systems and
Statistical Mechanics'', 3-13 July 2006. I am indebted to E.G.D. Cohen
for stimulating and clarifying comments on Eq.\equ(e5.2),\equ(e5.3).}

\*



\nota
\bibliographystyle{apsrev}

\revtex
\end{document}